\documentclass[aps,prl,twocolumn,showpacs,superscriptaddress,amsmath,amssymb]{revtex4-1}
\usepackage{graphicx}

\begin{document}
\title{Dislocations and the enhancement of superconductivity in odd-parity superconductor Sr$_2$RuO$_4$}

\author{Y. A. Ying}
\affiliation{Department of Physics and Materials Research Institute, The Pennsylvania State University, University Park, Pennsylvania 16802, USA}

\author{N. E. Staley}
\affiliation{Department of Physics and Materials Research Institute, The Pennsylvania State University, University Park, Pennsylvania 16802, USA}

\author{Y. Xin}
\affiliation{National High Magnetic Field Laboratory, Florida State University, Tallahassee, Florida 32306, USA}

\author{K. Sun}
\affiliation{Condensed Matter Theory Center and Joint Quantum Institute, Department of Physics, University of Maryland, College Park, Maryland 20742, USA}

\author{X. Cai}
\affiliation{Department of Physics and Materials Research Institute, The Pennsylvania State University, University Park, Pennsylvania 16802, USA}

\author{D. Fobes}
\affiliation{Department of Physics, Tulane University, New Orleans, Louisiana 70118, USA}

\author{T. Liu}
\affiliation{Department of Physics, Tulane University, New Orleans, Louisiana 70118, USA}

\author{Z. Q. Mao}
\affiliation{Department of Physics, Tulane University, New Orleans, Louisiana 70118, USA}

\author{Y. Liu}
\email{liu@phys.psu.edu}
\affiliation{Department of Physics and Materials Research Institute, The Pennsylvania State University, University Park, Pennsylvania 16802, USA}

\date{\today}

\begin{abstract}
We report observation of the enhancement of superconductivity near lattice dislocations and the absence of the strengthening of vortex pinning in odd-parity superconductor Sr$_2$RuO$_4$, both surprising results in direct contrast to the well known sensitivity of superconductivity in Sr$_2$RuO$_4$ to disorder. The enhanced superconductivity appears to be related fundamentally to the two-component nature of the superconducting order parameter, as revealed in our phenomenological theory taking into account the effect of symmetry reduction near a dislocation.
 \end{abstract}


\maketitle

Sr$_2$RuO$_4$, a leading candidate for a textbook example of unconventional superconductors\cite{MaenoNature1994,RiceJPCM1995,BaskaranPB1996,MackenzieRMP2003,MaenoJPSJ2012}, has attracted much attention in recent years in condensed matter physics and beyond, including the pursuit of non-Abelian Majorana anyons\cite{IvanovPRL2001,DasSarmaPRB2006}, key for topological quantum computing\cite{NayakRMP2008}. A large body of experimental data, including that obtained in phase sensitive measurements\cite{NelsonScience2004}, has shown that the layered perovskite material Sr$_2$RuO$_4$ is a spin-triplet, odd-parity superconductor\cite{MackenzieRMP2003,MaenoJPSJ2012}. Assuming that superconductivity in this material is two-dimensional (2D) in nature, the four-fold tetragonal crystalline symmetry dictates that the pairing symmetry in this superconductor must be one of the five representations\cite{RiceJPCM1995}. Among those, only the two-component, $p_x \pm  ip_y$ state is consistent with the muon spin rotation\cite{LukeNature1998} and Kerr rotation\cite{XiaPRL2006} measurements that suggest the presence of a spontaneous magnetic field in the superconducting state of Sr$_2$RuO$_4$, making it an electronic analog\cite{RiceJPCM1995} of the superfluid $^3$He A-phase and moreover a topological superconductor defined by the presence of gapless chiral edge states\cite{IvanovPRL2001,DasSarmaPRB2006}. An important question of interest is what is the microscopic mechanism responsible for such a highly exotic superconducting state. In this regard, models based on ferromagnetic\cite{MazinPRL1997} or antiferromagnetic fluctuations\cite{KuwabaraPRL2000}, spin-orbital coupling\cite{NgEPL2000}, interaction theory\cite{NomuraJPSJ2000,KoikegamiPRB2003}, and Hund's rule coupling\cite{BaskaranPB1996} have been proposed. The debate on these mechanisms is currently on going\cite{MaenoJPSJ2012}.

The eutectic phase of Sr$_2$RuO$_4$-Ru featuring crystalline islands of Ru embedded in the bulk crystalline Sr$_2$RuO$_4$, found previously to feature a $T_c$ nearly double that of the bulk Sr$_2$RuO$_4$\cite{MaenoPRL1998}, may provide insight into the mechanism issue. The $T_c$ enhancement was attributed to the capillary effect at the interface between Ru and Sr$_2$RuO$_4$\cite{SigristJPSJ2001}. It was more recently revealed that dislocations were abundant near a Ru island in bulk\cite{YingPRL2009}, also seen here in Sr$_2$RuO$_4$ flakes [Fig. 1(a)] (See Supplemental Material for detailed sample preparation), leading to the intriguing question as to whether these dislocations also contribute to the enhancement of $T_c$\cite{MackenziePRL1998}. An edge dislocation in a Sr$_2$RuO$_4$ lattice [Fig. 1(b)], near which the four-fold rotational symmetry is lost, is expected to lead to complicated modifications to the local lattice parameters and electronic states. 

On the other hand, a phenomenological theory can be formulated to capture the effect of the symmetry reduction associated with a dislocation without spelling out explicitly the local microscopic properties. The free energy density of the bulk Sr$_2$RuO$_4$ with a four-fold tetragonal symmetry in zero magnetic field can be written as\cite{SigristJPSJ2001},
\begin{equation}
\begin{split}
& f=a(|\eta_x|^2+|\eta_y|^2)+b_1(|\eta_x|^2+|\eta_y|^2)^2 \\
& +\frac{b_2}{2}(\eta_x^{\ast2}\eta_y^2+c.c.)+b_3|\eta_x|^2|\eta_y|^2 \\
& +K_1(|\partial_x\eta_x|^2+|\partial_y\eta_y|^2)+K_2(|\partial_x\eta_y|^2+|\partial_y\eta_x|^2) \\
& +[K_3(\partial_x\eta_x)^{\ast}(\partial_y\eta_y)+K_4(\partial_x\eta_y)^{\ast}(\partial_y\eta_x)+c.c.] \\
& +K_5(|\partial_z\eta_x|^2+|\partial_z\eta_y|^2)
\end{split}
\end{equation}
where $\eta_x$ and $\eta_y$ denote the two-component order parameter, $a(T) = \alpha(T-T_{c0})$, with $\alpha$ a constant and $T_{c0}$ = 1.5 K the bulk $T_c$ of Sr$_2$RuO$_4$, $b_i$ ($i = 1 - 3$) and $K_j$ ($j = 1 - 5$) parameters characterizing the bulk superconductor. Consider now a bulk Sr$_2$RuO$_4$ crystal for which the four-fold rotational symmetry is lost because of, say, the application of an in-plane uniaxial strain. A set of parameters, $m_1$ and $m_2$ used to quantify the effect of lattice distortions and $\mu$ to measure the mixing of the two order parameter components, can be introduced to describe the symmetry breaking strength. To obtain only the superconducting transition temperature, it is sufficient to consider the free energy density up to the quadratic terms,
\begin{equation}
\begin{split}
& f_{RS}=(a+m_1)|\eta_x|^2+(a+m_2)|\eta_y|^2 \\
& +\mu\eta_x^{\ast}\eta_y^{}+\mu^{\ast}\eta_x^{}\eta_y^{\ast}+\sum_{ijkl}\Gamma_{ijkl}\eta_i^{\ast}\eta_j^{\ast}\eta_k^{}\eta_l^{}
\end{split}
\end{equation}
The modified transition temperature $T_{ch}$, determined by the eigenvalues of the quadratic terms, is given by
\begin{equation}
T_{ch}=T_{c0}+\frac{1}{\alpha}\left (\sqrt{m_{-}^2+|\mu|^2}-m_{+}\right )
\end{equation}
where $m_{\pm}=(m_1 \pm m_2)/2$. Depending on the values of $m_{\pm}$, $T_{ch} > T_{c0}$ can be obtained [Fig. 1(c)]. 

\begin{figure}[!]
\includegraphics[viewport=100 120 510 670,scale=0.42]{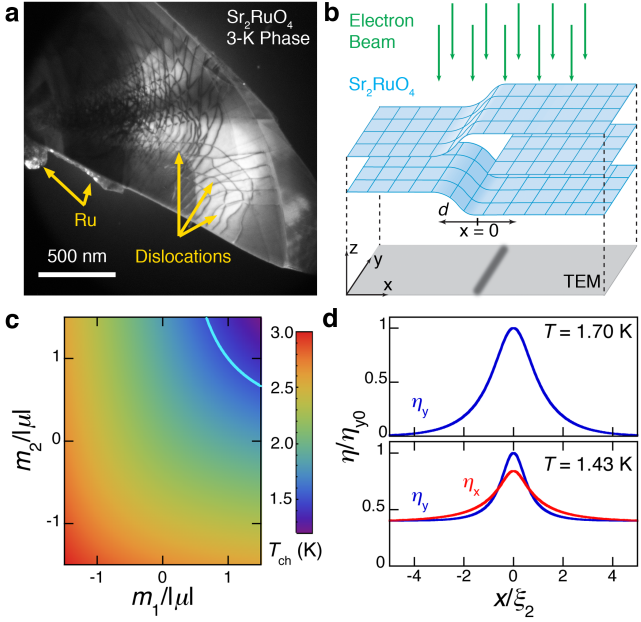}
\caption{(Color online) (a) Transmission electron microscopy (TEM) image of a Sr$_2$RuO$_4$ single-crystal flake showing Ru microdomains and dislocation lines. (b) Schematic of an edge dislocation caused by an extra layer in a Sr$_2$RuO$_4$ lattice. The edge dislocation scatters the electron beam in a TEM study, manifesting itself as a dark line in the TEM image. (c) $T_{ch}$ plotted as a function of $m_1/|\mu|$ and $m_2/|\mu|$ for $|\mu|/\alpha = 0.4$. The value of $T_{ch}$ is represented by a color scale. The highlighted curve represents the contour of $T_{ch}$ = 1.5 K. (d) Spatial dependence of $\eta_x$ and $\eta_y$ normalized to the value of $\eta_y$ at $x$ = 0 for $T_1$ = 1.7 K, $T_2$ = 1.43 K, and $m_1/|\mu|$ = -1.5, $m_2/|\mu|$ = -1.5, $|\mu|/\alpha$ = 0.4, respectively. Other parameters are $T_{c0}$ = 1.5 K, $T_{ch}$ = 3 K, $T_c$ = 1.9 K, and $\xi_2(T_{1,2}) = [K_2/|\alpha(T_{1,2} -T_{c0})|]^{1/2}$, superconducting coherence lengths at finite temperatures.} 
\label{Fig1}
\end{figure}

Building on this result for a hypothetical bulk crystal of Sr$_2$RuO$_4$ with a broken four-fold symmetry, we proceed to analyze a more realistic model system featuring a single dislocation of a width $d$ embedded in a bulk crystal, shown in Fig. 1(b). To begin with, we assume that the locally enhanced superconducting transition temperature within the dislocation is $T_c$, which is lower than $T_{ch}$ for the corresponding homogeneous bulk characterized by a set of the symmetry reduction parameters ($m_1$, $m_2$, and $\mu$). To reach $T_c$, the effect of the bulk on the embedded dislocation region must be taken into account. This can be done by writing the free energy density for the dislocation region in the form of a $\delta$-function
\begin{equation}
f_D=\delta(x)d\alpha(T-T_{c0})(|\eta_x|^2+|\eta_y|^2)
\end{equation}
employed successfully in the previous analysis of the capillary effect at the interface between Ru and Sr$_2$RuO$_4$\cite{SigristJPSJ2001}, and solving the linearized Ginzburg-Landau equations derived from Eq. (1). Matching the boundary conditions at $x$ = 0 (see Supplemental Material), we found that $T_c$ is given by the solution of
\begin{equation}
2\sqrt{\frac{K_2}{\alpha}(T_c-T_{c0})}=d(T_{ch}-T_c)
\end{equation}
which requires that $T_c > T_{c0}$ and $T_{ch} > T_c$ for simple self consistency. Because the highest $T_c$ found in Sr$_2$RuO$_4$ and the related eutectic systems is 3 K\cite{MaenoJPSJ2012,MaenoPRL1998}, we assume that $T_{ch}$ = 3 K. As a result, the experimentally observed $T_c$ of 1.9 K for Sample 1 (see below) corresponds to $d = 1.41(K_2/\alpha T_{c0})^{1/2}$. This $d$ value is comparable to the superconducting coherence length given by $\xi_{1,2}(0) = (K_{1,2}/\alpha T_{c0})^{1/2}$ for the anisotropic stripe, suggesting that the use of a $\delta$-function is reasonably self-consistent given that the basic length for order parameter variation in the Ginzburg-Landau theory is $\xi_{1,2}(0)$.

The $y$-component of the order parameter, $\eta_y$, was found to first become non-zero for $T < T_c$ as shown in Fig. 1(d) (upper panel), making the dislocation a nucleation center for superconductivity. Below $T_{c0}$, both $\eta_x$ and $\eta_y$ are present, resulting in an enhanced order parameter near the dislocation [Fig. 1(d), lower panel]. The latter suggests that an edge dislocation in Sr$_2$RuO$_4$ is not a pinning center for Abrikosov vortices (see below) as opposed to that in high-$T_c$ superconductors\cite{SandiumengeAM2000}. It is interesting to note that similar phenomenology can be obtained if two pairing states represented by a single-component order parameter have identical or very close intrinsic superconducting transition temperatures.

\begin{figure}[!]
\includegraphics[viewport=100 260 510 540, scale=0.42]{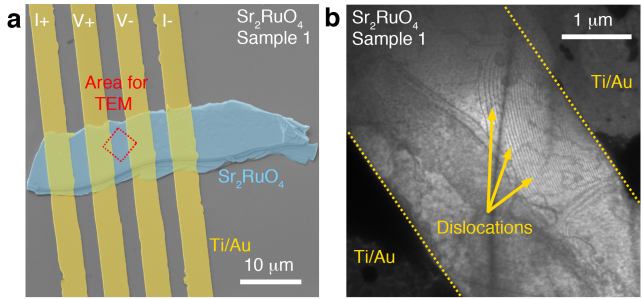}
\caption{(Color online) (a) False-color scanning electron microscopy (SEM) image of Sample 1. The red square indicates the area examined by TEM. (b) TEM image of the boxed area in (a), showing dislocation lines but no Ru microdomains. The region not shown in this image was also checked by SEM and TEM and found to possess no Ru microdomains.}
\label{Fig2}
\end{figure}

Experimentally, easily cleavable single crystals of Sr$_2$RuO$_4$ were synthesized by the floating-zone method. The bulk $T_c$ was found to be 1.35 K without the presence of enhanced superconductivity (Supplemental Material Fig. 1), suggesting that very few, if any, Ru microdomains were present in our starting crystals. To overcome the unavailability of superconducting films of Sr$_2$RuO$_4$ in spite of an early report of initial synthesis success\cite{KrockenbergerAPL2010}, we prepared single-crystal flakes with a lateral dimension of roughly $10 - 50$ $\mu$m and a thickness of $300 - 800$ nm by mechanical exfoliation. The flakes were transferred onto a Si/SiO$_2$ substrate with the $c$ axis of the crystal perpendicular to the substrate. A standard four-point pattern was defined on the crystal via contact photolithography, followed by a 30 s oxygen plasma operated at 100 mTorr and 100 W to remove any residue photoresist before metallization. Electrical leads of 50 nm Ti and 200 nm Au were then deposited at 45-degree angles with respect to the substrate norm to ensure the continuity of the leads on the side walls of the crystals, resulting in contact resistances less than 1 $\Omega$ at low temperatures. Low temperature dc measurements were performed in a $^3$He refrigerator with a base temperature of 0.35 K. The critical current was defined by the bias current corresponding to a voltage response of 50 nV.

Single-crystal flakes of Sr$_2$RuO$_4$ were examined by scanning electron microscopy (SEM) and transmission electron microscopy (TEM). Both dislocations and Ru microdomains were found in some flakes with Ru microdomains always locating at the edge of the crystal [Fig. 1(a)], perhaps because the cleaving tends to occur at the Ru/Sr$_2$RuO$_4$ boundary due to different mechanical strengths of Ru and Sr$_2$RuO$_4$. For some flakes, however, no traces of Ru microdomains were found. In particular, in a four-wire Sr$_2$RuO$_4$ device, Sample 1, no Ru microdomains were found in the SEM [Fig. 2(a)] and the TEM [Fig. 2(b)] images taken before and after the low-temperature measurements, respectively. Meanwhile, dislocations were observed between the two voltage leads. Superconductivity with an enhanced onset $T_c$ = 1.9 K was found [Fig. 3(a)], directly linking the presence of dislocations and an enhancement of superconductivity in Sr$_2$RuO$_4$. A set of symmetry reduction parameters of $m_1/|\mu| = -1.5$, $m_2/|\mu| = -1.5$, $|\mu|/\alpha = 0.4$, and $d/\xi_2(0) = 1.41$ characterizing the dislocation will give rise to $T_{ch}$ = 3 K and $T_c$ = 1.9 K in the phenomenological theory presented above.

\begin{figure}[!]
\includegraphics[viewport=80 50 500 680,scale=0.52]{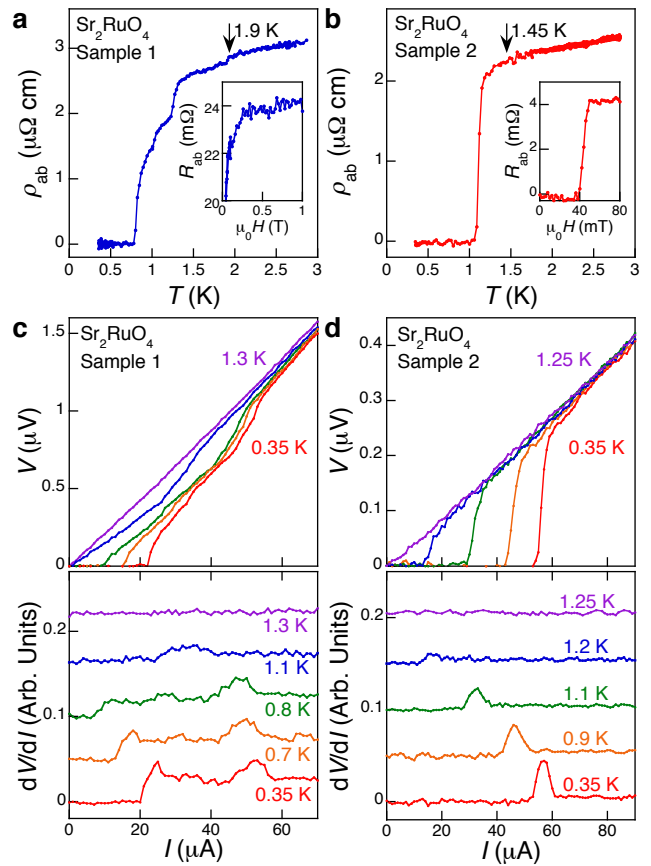}
\caption{(Color online) In-plane resistivity $\rho_{ab}$ of (a) Sample 1 and (b) Sample 2 taken at zero applied magnetic field, plotted as a function of temperature. Insets: Magnetic field dependence of $\rho_{ab}$ at 0.35 K, for field applied along the $c$ axis. The upper critical field of Sample 1 was found to be about 0.3 T, suggesting the presence of enhanced superconductivity. Zero-field $V-I$ curves and corresponding $dV/dI-I$ curves at various temperatures for (c) Sample 1 showing multiple transitions and (d) Sample 2 showing a single transition. The $dV/dI-I$ curves except for those at 0.35 K were shifted for clarity.}
\label{Fig3}
\end{figure}

The presence of multiple dislocations [Fig. 2(b)], which should be described by different sets of symmetry reduction parameters, suggests that multiple phases may be present in Sample 1. This is consistent with the multiple features observed in the $R(T)$ curve [Fig. 3(a)]. The voltage-current ($V-I$) characteristics and the $dV/dI-I$ curves were found to show double features suggesting the existence of two different phases at low temperatures [Fig. 3(c)], one corresponding to the dislocations and the other the bulk phase. In contrast, in Sample 2 (see SI Fig. 3 for the SEM image), a single onset $T_c$ slightly lower than 1.5 K [Fig. 3(b)] and a single feature [Fig. 3(d)] were found in the $R(T)$, $V-I$, and $dV/dI-I$ curves, respectively. 

The presence of dislocations in a type II superconductor is expected in general to enlarge the critical current density ($J_c$), which measures the strength of the vortex pinning, because Abrikosov vortices tend to be pinned to structural defects\cite{BlatterRMP1994}. Essentially, the energy cost of placing a normal vortex core at a defect is in general smaller than the defect-free part of the sample, making it a preferred site for vortex pinning. Strengthened vortex pinning due to dislocations was indeed reported in high-$T_c$ superconductors\cite{SandiumengeAM2000}. For Sr$_2$RuO$_4$, on the other hand, $J_c$ would not be enhanced even when dislocations are present in the sample because a dislocation gives rise to locally enhanced rather than suppressed superconductivity. We would therefore expect samples with or without an enhanced $T_c$ to show similar $J_c$ values. 

\begin{figure}[!]
\includegraphics[viewport=80 90 500 770,scale=0.51]{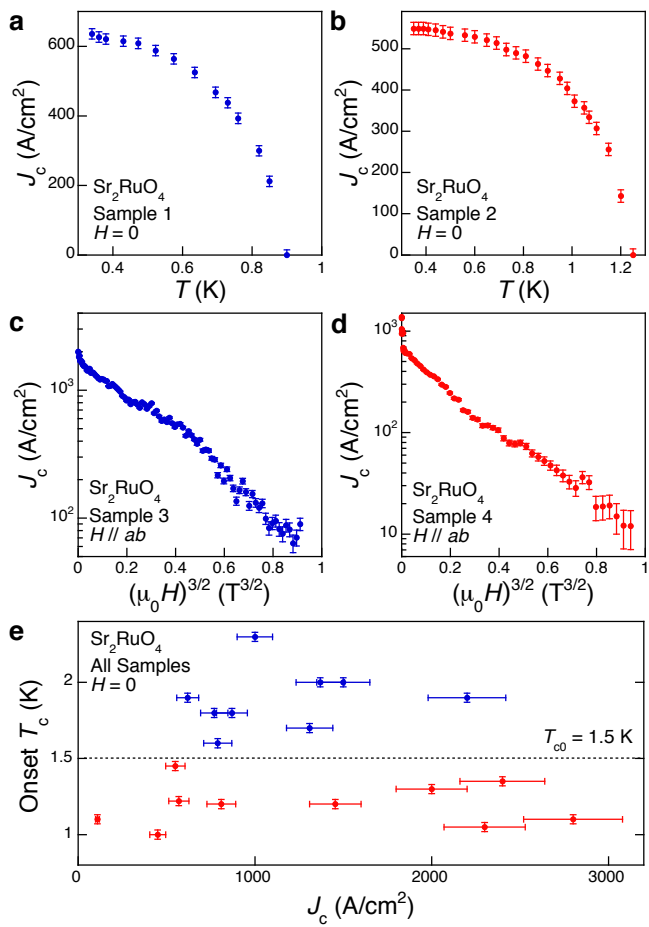}
\caption{(Color online) Zero-field critical current density $J_c$ plotted as a function of temperature for (a) Sample 1 and (b) Sample 2. Base temperature (0.35 K) $J_c$ plotted logarithmically as a function of $H^{3/2}$ for (c) Sample 3 and (d) Sample 4, showing roughly straight lines, consistent with the prediction of collective pinning theory. The magnetic field was applied parallel to the in-plane direction along some of the dislocation lines. (e) Onset $T_c$ for samples with an enhanced $T_c$ (Blue) and pure phase samples (Red) plotted as a function of their base temperature $J_c$ at zero-field.}
\label{Fig4}
\end{figure}

We measured $J_c$ of our Sr$_2$RuO$_4$ flakes with an in-plane current (Fig. 4). Values of $J_c$ were found to be much smaller than the depairing current density at the thermodynamic limit ($J_0 = cH_c/3\sqrt{6}\pi\lambda_{ab} = 5\times10^6$ A/cm$^2$, where $H_c$ = 20 mT is the thermodynamic critical field and $\lambda_{ab}$ = 190 nm the in-plane penetration depth), but larger than $J_c$ values with the current applied along the $c$ axis measured previously in bulk Sr$_2$RuO$_4$\cite{HooperPRB2004,HooperPRB2006}. As expected, the temperature and magnetic field dependences of $J_c$ were found to be similar in samples with and without an enhanced $T_c$, showing that the presence of dislocations does not enhance vortex pinning. In addition, it was found that $J_c/J_0 \sim 10^{-4} - 10^{-3}$, indicating that the pinning potential is weak\cite{BlatterRMP1994}, and furthermore, $J_c(H) \sim exp[-(H/H_0)^{3/2}]$ in the intermediate fields, consistent with that predicted by the collective pinning theory of vortex lattices\cite{BlatterRMP1994}. The latter suggests that the vortices are pinned collectively by point defects instead of dislocations. Measurements on a large set of samples showed $J_c$ values to fall into the same range regardless of whether a $T_c$ enhancement was found in the sample [Fig. 4(e)], showing convincingly that dislocations are not pinning centers in Sr$_2$RuO$_4$, as predicted by our phenomenological theory. 

Useful insight into the mechanism of odd-parity superconductivity in Sr$_2$RuO$_4$ can be obtained by examining the microscopic origins of the $T_c$ enhancement associated with a dislocation. An edge dislocation, as well as a screw dislocation, destroys locally both the four-fold rotational symmetry and layering along the $c$ axis [Fig. 1(b)]. The topological nature of a dislocation demands that a number of the adjacent layers be placed locally closer than those in the bulk\cite{RanNP2009}. The local electronic states are then strongly restructured, becoming at the same time more three-dimensional (3D). Interestingly, the quantum oscillation measurements carried out under a hydrostatic pressure\cite{ForsythePRL2002} suggest that the pressure, which lowers $T_c$ of Sr$_2$RuO$_4$, makes the Fermi surface more 2D like. Furthermore, applying a uniaxial pressure along the $c$ axis of Sr$_2$RuO$_4$, which should increase the interlayer coupling, was found to enhance $T_c$\cite{KittakaPRB2010}. All these observations suggest that an enhanced $T_c$ in Sr$_2$RuO$_4$ may originate from the strengthening of the interlayer scattering of electrons near a dislocation, which may also lead to a strongly $p_z$-dependent order parameter. Such a strongly $p_z$-dependent order parameter was obtained in a band-dependent superconductivity model for Sr$_2$RuO$_4$\cite{ZhitomirskyPRL2001}, which, incidentally, results in a strongly suppressed depairing current\cite{KeePRB2004}, explaining naturally the small $J_c$ values observed in Sr$_2$RuO$_4$.

The authors acknowledge useful discussions with Y. Maeno, C. C. Tsuei, S. B. Chung and V. Varkaruk. The work at Penn State is supported by DOE under Grant No. DE-FG02-04ER46159. The nanofabrication part of the work is supported by Penn State MRI Nanofabrication Lab under NSF Cooperative Agreement 0335765, NNIN with Cornell University and under NSF DMR 0908700. The TEM images were obtained at the TEM facility at FSU, supported by the Florida State University Research Foundation, NSF-DMR-0654118 and the State of Florida. K. S. is supported by JQI-NSF-PFC. The work at Tulane is supported by NSF under DMR-0645305.

\end{document}